# Orchestrating optical pumping with magnetic field for enhanced magnetometry

SWARUPANANDA PRADHAN,[1,2,*] SUDIP MANDAL[1]

[1]*Photonics and Nano Technology Section, Atomic and Molecular Physics Division, Bhabha Atomic Research Centre Facility, Visakhapatnam-531011, India*

[2]*Homi Bhabha National Institute, Department of Atomic Energy, Mumbai-400094, India*

*Corresponding author: spradhan@barc.gov.in*

**The atomic population trapped in uncoupled atomic states is a limiting factor for processes based on laser-atom interaction. The use of repump laser, bi-chromatic field, and vector magnetic field are explored in degenerate as well as non-degenerate atomic system to overcome the limitation. The magnetic resonance of $^{85}$Rb atoms under these complementary conditions are investigated in a buffer gas filled cell. The most suitable condition is achieved by the combined action of a bi-chromatic light with a specific combination of magnetic fields for non-degenerate levels where the relevant magnetic resonance shows more than 7-fold increase in its amplitude. The phenomenon is consistently observed through RF as well as magnetic field de-modulation technique. It will lead to similar improvement in the sensitivity of the atomic magnetometers. The optical pumping of the atomic population to the stretch state by manipulation of magnetic field will be of general interest involving quantum technology.**

## 1. Introduction

The precise control of microscopic to macroscopic properties through magnetic field has broader applications like inducing tuneable inter-atomic interaction across Feshbach resonance [1], regulating structure of atomic wires [2], quantum information processing [3], Sisyphus cooling [4], quantum sensors [5, 6] etc. Conversely, coherent manipulation and reading of the quantum dynamics of atoms in a magnetic field is key to atomic magnetometry [7-10]. There has been remarkable progress in the techniques for magnetic field sensing due to its widespread application. The atomic magnetometry is emerging as a frontrunner among contemporary techniques due to its operation near room temperature, ultra-high sensitivity, advance functionality and flexible configuration. The improved knowledge of the underlying processes behind the magnetic resonances is vital to extend these limits to unprecedented level.

The optical pumping to the uncoupled atomic hyperfine level is a primary cause for reduction of the optical density in atomic cell [11]. It is extreme in buffer gas filled cell, that is popularly used for a variety of quantum sensor. The sensitivity of these atomic sensors will be significantly improved by utilizing the lost atomic population. In this article, we explore methods to circumvent it for improved sensitivity of the atomic magnetometer. The various ways to pump back (repump) from the inaccessible level are envisaged. The attributes of the magnetic resonance are the basis of magnetometer and provides clues for underlying physical processes. The magnetometry geometry is another critical parameter that dictates compact operation of the device. The inter-comparison among various methods is carried out based on these factors. The first method involves use of an additional laser beam to bring the lost population trapped in the uncoupled hyperfine level. The requirement of additional laser beam is avoided by using a bi-chromatic field generated from a single laser source. The most favorable result is obtained by using a bi-chromatic light field in combination with longitudinal and orthogonal magnetic field. The phenomenal increase in the magnetic resonance under this combined action is established by using two complementary signal acquisition technique and is suitable for compact operation of the magnetometer.

## 2. Experimental apparatus:

The experiment is performed on Rb atoms (natural isotopic composition) with $N_2$ gas @ 25 Torr. The schematic of the experimental diagram (Fig.-1) shows a vertical cavity surface emitting laser (VCSEL) @795 nm output interacting with atomic sample kept in a temperature and magnetic controlled enclosure, and is termed as the probe laser. A small part of the beam, tapped through a polarization beam splitter cube (PBS) PBS1 is used for stabilizing laser frequency using absorption spectroscopy. The frequency of external cavity diode laser (ECDL) @780 nm is stabilized with reference to a saturation absorption spectroscopy (SAS) spectrum. The ECDL beam is used for recycling the atoms lost to the 5 $^2S_{1/2}$ F=2 level and is termed as the repump laser.

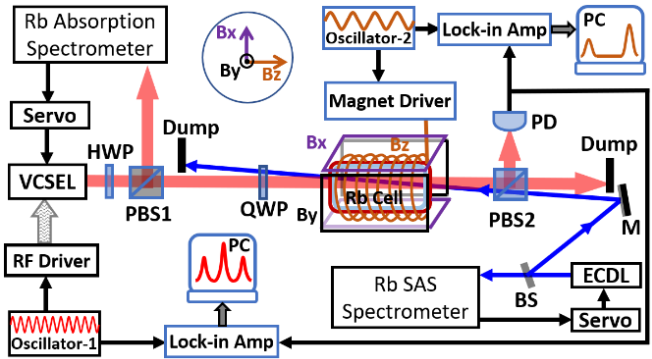

Fig.-1: Schematic diagram showing essential components of the experimental set-up. The frequency of the VCSEL and ECDL are frequency stabilized to Rb spectrum. The Rb cell contains $N_2$ gas at 25 Torr and is kept in a fixed temperature and magnetic field enclosure. The RF modulates the laser current near 1.517 GHz to generate the side bands. The transmission of the VCSEL beam through the atomic cell is studied with respect to the ECDL beam, RF, and magnetic field.

The experiments are carried out with degenerate energy levels (DL) as well as non-degenerate energy levels (NDL) to address the detrimental effect of optical pumping. For experiments involving DL, the reflected light intensity across the analyser PBS (PBS2) is detected by a lock-in amplifier with respect to the Bz magnetic field modulation (Freq. 55 Hz, Amp. 160 nT). The Bz field is scanned across the zero-magnetic field to acquire the magnetic resonance. The quarter waveplate (QWP) angle is kept at $+20^0$ where the reflected polarization rotation signal is nearly maximized [10, 12, 13]. The influence of repump laser as well as probe laser side bands (at 1.517GHz) on the magnetic resonance is studied. The magnetic field along x and y direction are kept close to zero value.

For experiments involving NDL, the photo diode (PD) is placed at the location of PBS2 in Fig.-1. The PD signal is processed through a lock-in amplifier triggered to the modulation applied to the Bz Field with above modulation parameters or modulation applied to the RF (Freq. 440 Hz, Amp. 2 kHz). These demodulation techniques provide complementary information. The Zeeman degeneracy is lifted by applying a longitudinal Bz magnetic field + 5 µT. The radio frequency (RF) is scanned across the two-photon resonance (1.517866 GHz) with QWP angle at $45^0$. The influence of orthogonal By magnetic field (~7.5 µT) on the magnetic resonance is studied.

## 3. Results and Discussion:

The optical pumping to the uncoupled atomic state can be prevented by working at lower light intensity, using a repump laser to pump back atomic population, or use of multiple light frequency generated from a single laser source. The lowering of light intensity will lead to compromise in the quality of the signal of interest in many situation and is not a generic solution. We explore the influence of a repump light on the amplitude of the zero-field magnetic resonance of $^{85}$Rb atoms as shown in Fig.-2. It uses DL configuration and measures the reflected signal through the PBS. The repump laser being tuned to Rb-D2 transition, does not share any common level with probe beam coupling channel and solely performs the repumping action. It pumps most of the population in the F=2 level to F=3 ground level [11]. Consequently, there is nearly 6-fold increase in the amplitude of the magnetic resonance. The magnetometer will exhibit similar increase in the sensitivity with the use of an additional repump laser beam.

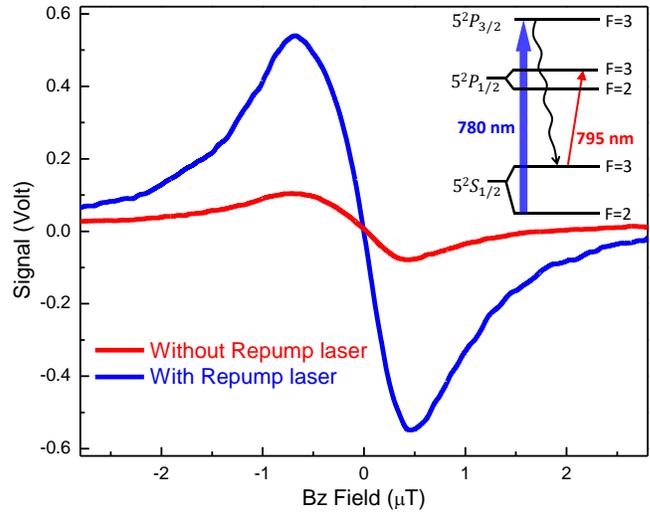

**Fig.-2:** Zero-field magnetic resonance without (red) and with (blue) control laser tuned to 780 nm $^{85}$Rb D2 transition. The probe laser is locked to the 795 $^{85}$Rb D1 F=3→F'=3 transition. The cell temperature was kept at~22 $^0$C. The angle of the QWP was kept at ~$20^0$ in the path of probe beam, prior to the atomic cell. (insert) Energy level showing coupling of the repump and probe laser light. There is no common level in the coupling of probe and repump light.

The complexity of an additional laser beam in magnetometry is avoided by using frequency modulated light field. The VCSEL current is modulated to generated sidebands that are coupled with both lower hyperfine level, thereby assessing the whole atomic population. However, it comes at an expense of introducing competing physical processes as both bright and dark resonances contribute to the signal for elliptically polarized light field (QWP = $20^0$). The $F_g \rightarrow F_e \leq F_g$ and $F_g \rightarrow F_e > F_g$ class of transition gives rise to dark (enhanced transmission) and bright (enhanced absorption) resonances respectively [10, 14-17]. The $F_g = 2 \rightarrow F_e = 3$ is a bright resonance, whereas the other three possible transitions are dark resonance. Consequently, an enhanced transmission (positive slope) is observed for the frequency modulated light as shown in Fig.-3. However, the extent of increase in the signal amplitude is merely ~1.4 times as compared to ~6-fold increase in presence of repump laser beam. The minuscule increase despite assessing the whole atomic population is due to the competing nature of the resonances.

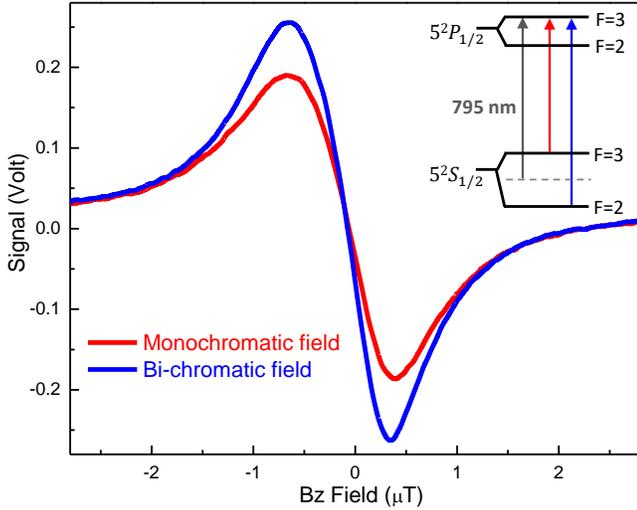

**Fig.-3:** Zero-field magnetic resonance with monochromatic (red) and bi-chromatic laser field (blue). The latter is generated by modulating the laser frequency at 1.5178 GHz. The cell temp is kept at ~22 °C and QWP angle in the path of probe laser beam was at ~20°. (insert): coupling of the side bands (bi-chromatic light) with the Rb energy level is shown. The side band are kept away from two-photon Raman resonance to study the role of optical pumping.

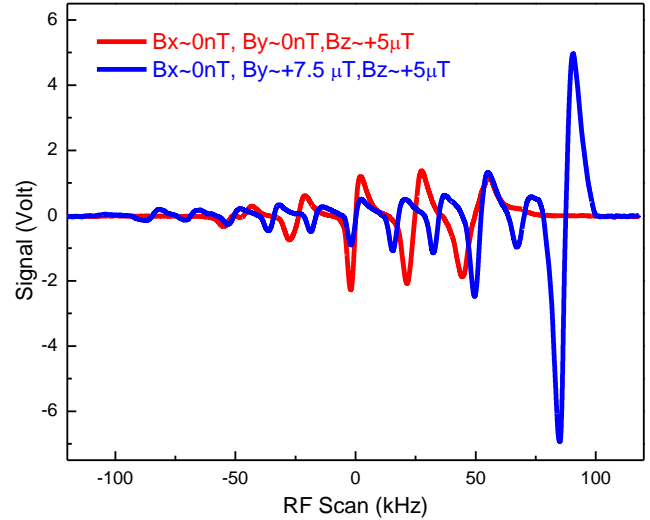

**Fig.-4:** CPT resonance through demodulation with respect to modulation given to RF field. The zero of the RF scanning corresponds to 1.517866 GHz. The Bx and Bz field are at ~0 nT and ~+5 µT. The red and blue curves are taken at By ~ 0 nT and ~+7.5 µT. The increase in the amplitude of the signal corresponding to stretch state in blue curve shows enhanced optical pumping due to the orthogonal By field.

The use of bi-chromatic field for DL configuration brought additional complexity in magnetometry that diminished the advantage of assessing trapped population. We explore the possibility of improving amplitude of the magnetic resonance by using bi-chromatic field in NDL configuration. The degeneracy is lifted by applying a Bz magnetic field of ~ + 5 µT. The bi-chromatic field is coupled with both hyperfine ground levels. So, the difficulty of inaccessible states and opposite characteristic (bright and dark resonance in DL) of transition are simultaneously resolved. The experiments are carried out by scanning the RF field around the two-photon resonance (1.517866 GHz). The transmitted light intensity is demodulated with respect to the modulation applied to the RF (RFM) or Bz magnetic field (MM). The transmission of the circularly polarized light through the atomic cell (without PBS2) provides the coherent population trapping (CPT) signal. Since the signal is acquired as a function of RF field, the bias magnetic field can be applied in any direction.

The bias Bz field splits the CPT resonances as shown in Fig.-4. The resonances are well studied in absence of any orthogonal fields in past literature. In presence of a bias By (or Bx) field, additional resonances appears due to change in the quantization axis that alters the selection rule [18]. We observe phenomenal increase in the amplitude of the stretched resonance as the By magnetic field is increased. It attains an optimum value for By~+7.5 µT. The increase in amplitude is >7 times compared to that in absence of orthogonal field. The decrease in the amplitude of the rest of the resonances is a clear indication of enhanced Zeeman optical pumping by the orthogonal field to the stretch state. It demonstrates a convenient method to play with optical pumping for improved magnetometry without compromising on the size of the device. Importantly, it opens an interesting window to control the optical pumping by manipulating magnetic field.

The CPT signal through demodulation of the light intensity with respect to the magnetic field modulation provides a complementary way for magnetometry. Here, only the magnetic field sensitive resonances are observed. Thus, the resonance corresponding to (mf=0, mf=0) coupling is not observed. The stretched CPT resonance shows remarkable increase in its amplitude for By magnetic field ~+7.5 µT. The similar observation in Fig.-4 and Fig.-5 proves the consistency of the phenomenon. The increase in the level of the signal amplitude for RF demodulation and magnetic field demodulation is in the same range.

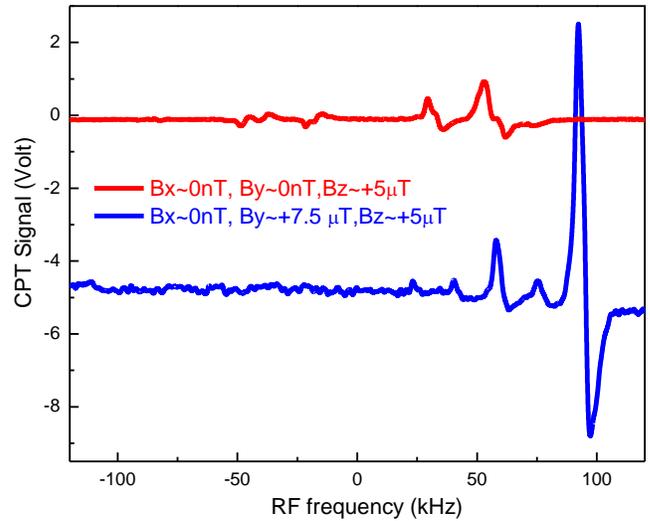

**Fig.-5:** CPT resonance through demodulation with respect to modulation given to Bz magnetic field. The Bx and Bz field are at ~0 nT and ~+5 µT. The red and blue curves are taken at By ~ 0 nT and ~+7.5 µT. The increase in the amplitude of the stretch state signal in blue curve is consistent with Fig.-4.

There is a change in the DC background level of the signal with change in the bias By field in Fig.-5. The Bloch equation $\boldsymbol{B} \times \boldsymbol{P} + P_0 \Delta B = \Delta B\, \boldsymbol{P}$ is a simple way to describe the dynamics of spin polarization $\boldsymbol{P}$ in a magnetic field $\boldsymbol{B}$ under steady state condition with $P_0$ representing the equilibrium spin polarization and $\Delta B$ to the associated population transfer rates [19]. The phase sensitively detected signal with respect to Bz modulation corresponds to the partial derivative (with respect to Bz) of the atomic polarization along the laser beam (Pz). The solution of the Bloch equation gives

$$\frac{\partial P_z}{\partial z} \sim 2 P_0\, B_z\, \frac{B_x^2 + B_y^2}{\left(B_x^2 + B_y^2 + B_z^2 + \Delta B\right)^2} \qquad [1]$$

The increase in the By field will lead to the increase in the amplitude of the magnetic field demodulated signal as given in Eq.-1. It suitably explains the DC offset of the signal for By~+7.5 μT in Fig.-5.

The increase in the amplitude of the stretched signal ceases for larger Bz field (>~+6 μT) with By fixed at +7.5 μT. The amplitude of the resonances approaches to that in absence of orthogonal field at higher value of Bz Field. Wynands *etal* has used simplified model to analyse the strength of the CPT resonances [18]. It is shown that the optical pumping plays a dominating role in conjunction with the strength of the resonance. The buffer gas is generally believed to be reducing the optical pumping due to increase in the homogeneous linewidth and depolarization of the excited state. The confinement induced enhanced optical pumping is a competing process in buffer gas environment that has been recently investigated [11]. The longer confinement of the atoms in the light field and mixing of the state's due to orthogonal field is responsible for the remarkable enhancement of the stretched signal for a specific combination of orthogonal and longitudinal magnetic field in buffer gas filled environment [11,18]. The analysis of the experimental observation with the Lindblad master equation will provide a comprehensive picture of the process [15, 20, 21].

## 4. Conclusions:

The complementary methods to use inaccessible atomic population for magnetometry application are explored. The use of a repump laser shows ~ 6-fold increase in the strength of the magnetic resonance as compared to ~1.4 times for the use of a bi-chromatic field. The addition of a longitudinal and orthogonal magnetic field in bi-chromatic field configuration shows remarkable (> 7-fold) increase in the strength of the magnetic resonance. It overcomes the problem of opposite polarity of the coupled magnetic resonances while accessing the whole atomic population. It provides an interesting approach to transfer population to the stretch state by playing with the magnetic field (without increase in the light intensity). This method will be suitable for compact operation of the magnetometer in contrast to the use of an additional repump beam. The fundamental of the associated physical processes will be useful in quantum technology relying on coherent manipulation of atomic dynamics using light and magnetic field.

**Acknowledgements.** The authors are thankful to D. Rao for going through the manuscript and providing important suggestions. The authors are thankful to D. Udupa, S. M. Yusuf and A. K. Mohanty for supporting the activity.

**Disclosures.** The authors declare no conflicts of interest.

**Data availability.** Data underlying the results presented in this paper are not publicly available at this time but may be obtained from the authors upon reasonable request.